\documentclass[12pt]{article}
\usepackage{times}
\usepackage{psfig}
\usepackage{multicol}
\usepackage{wrapfig}
\usepackage{graphicx}

\newcommand{\Msun}      {\mbox{$\rm\,M_{\mathord\odot}$}}

\begin{document}

\begin{figure}[t]
\vspace{-2.5cm}
\hspace{-1.7cm}
\psfig{figure=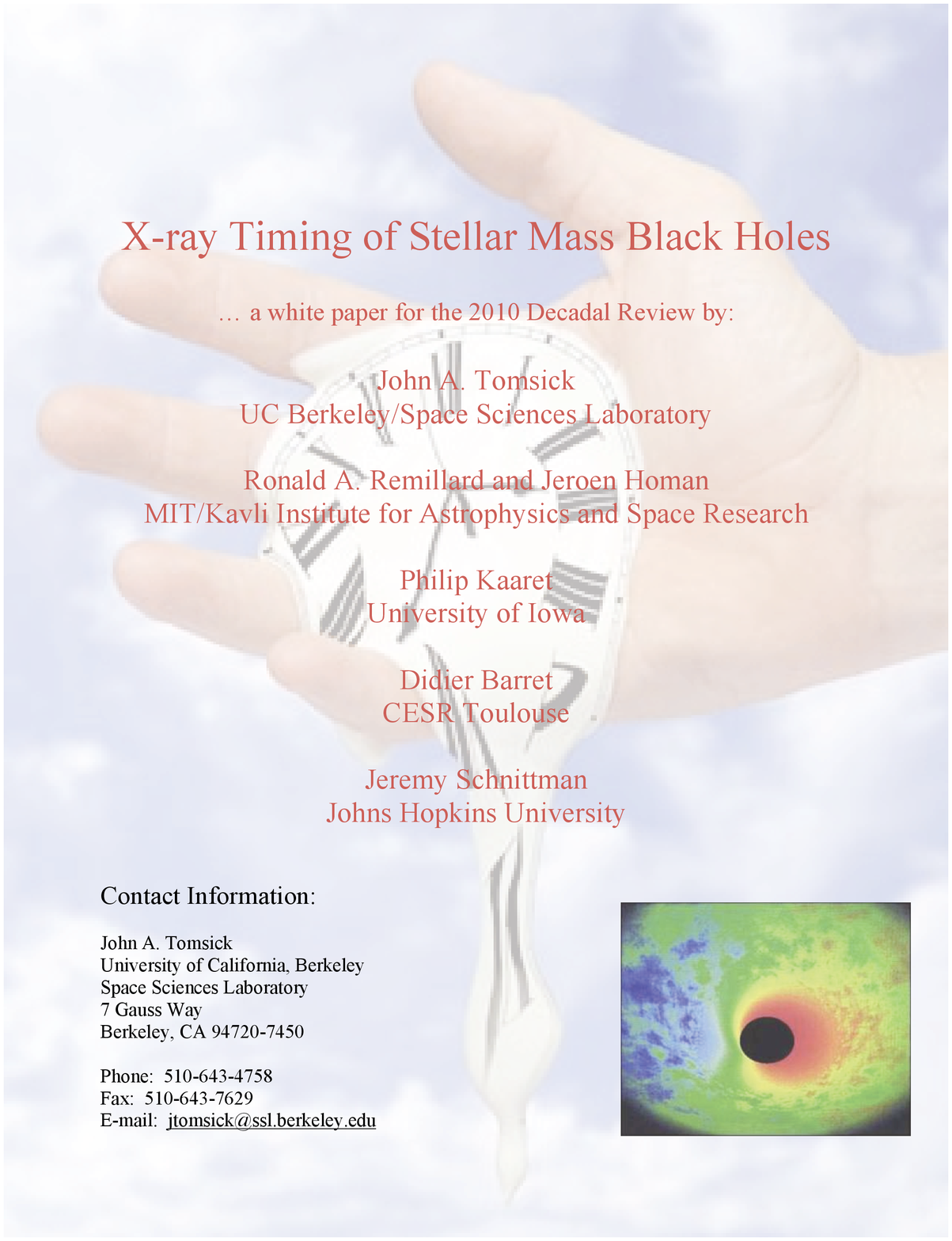,height=26.1cm,angle=0}
\end{figure}

\clearpage

\section{Abstract} 

X-ray timing observations of accreting stellar mass black holes have 
shown that they can produce signals with such short time scales that 
we must be probing very close to the innermost stable circular orbit
that is predicted by the theory of General Relativity (GR).  These 
signals are quasi-periodic oscillations (QPOs), and both the 
high-frequency variety (HFQPOs, which have frequencies in the 
40--450~Hz range) as well as the 0.1--10~Hz low-frequency type 
have the potential to provide tests of GR in the strong field limit.
An important step on the path to GR tests is to constrain the physical
black hole properties, and the straightforward frequency measurements 
that are possible with X-ray timing may provide one of the cleanest 
measurements of black hole spins.  While current X-ray satellites 
have uncovered these phenomenona, the HFQPOs are weak signals, and
future X-ray timing missions with larger effective area are required
for testing the candidate theoretical QPO mechanisms.
Another main goal in the study of accreting black holes is to 
understand the production of relativistic jets.  Here, we have also
made progress during the past decade by finding clear connections
between the radio emission that traces the strength of the jet and
the properties of the X-ray emission.  With new radio capabilities
just coming on-line, continuing detailed X-ray studies of 
accreting black holes is crucial for continuing to make progress.

\section{Introduction and Scientific Goals}

Studies of accreting stellar mass black holes address some of the 
most significant questions in astrophysics, including probing 
regions of strong gravity close to the black hole as well as 
improving our understanding of ubiquitous phenomena such as 
accretion disks and jets.  The few dozen black holes (BHs) or 
black hole candidates (BHCs) that we have found in the Galaxy 
\cite{rm06} show X-ray variability on time scales from 
milliseconds to years as they accrete matter from their binary 
stellar companions.  Their X-ray fluxes can vary by factors of 
more than a billion, becoming, at times, the brightest sources 
in the X-ray sky.

That these sources have great potential for helping us to learn
about fundamental physics has not gone unnoticed, and accreting 
BHs received significant attention in the 2000 Decadal Review.
The panel on High-Energy Astrophysics from Space included BH
science in forming its list of ``astrophysics challenges''
\cite{blandford00}.  Specifically, these include the challenge to 
``form an indirect image of the flow of gas around a black hole'' 
and to ``understand how jets are created and collimated.''
With X-ray timing at the millisecond level, we are probing the
regions of the accretion flow at the very inner edge of the disk.
Not only do we know that there is variability coming from these
regions, but there are signals at specific frequencies 
(quasi-periodic oscillations or QPOs) that provide quantitative 
constraints on the properties of the inner disk.

In the following, we first discuss the major progress that has
been made using X-ray timing to study BHs over the past decade.  
We then focus on opportunities X-ray timing can provide in the
next decade related to the goals of testing General Relativity 
(GR) in the strong field limit and understanding jets.  Finally,
we discuss the observational requirements that an X-ray timing 
mission would need to take advantage of these opportunities.

\section{Black Holes in 2010:  Discoveries and Progress to Date}

{\bf X-ray binaries} provide the best opportunities for constraining 
the {\bf mass} of stellar-mass BHs. In the past decade, almost 20
confirmed and candidate BHs were observed in outburst, including 10 
new discoveries. Eleven new sources were added to the list of 
dynamically confirmed BHs, doubling the total number.  The dependence 
of X-ray timing properties on BH mass is clearly demonstrated by 
comparing the timing properties of stellar-mass BHs and the BHs in 
AGN.  Using the break frequencies observed in BH power density spectra 
(PDSs), it has been shown that, after correcting for mass accretion 
rate, AGN behave like scaled-up stellar-mass BHs \cite{mchardy06}.  
Similarly, QPOs have been used to argue that at least some 
Ultra-Luminous X-ray sources (ULXs) harbor intermediate-mass black 
holes \cite{casella08}.

Dedicated monitoring of transient sources, in which their evolution 
over periods of weeks to months is followed on a $\sim$daily basis, 
have been critical in improving our understanding of the correlated 
behavior of X-ray timing and spectral properties, i.e., the 
{\bf black-hole X-ray states}. This has led to a clear organization 
of BH behavior into three active states \cite{rm06}. Each state 
displays different combinations of energy spectra, PDSs, and 
multi-wavelength properties, implying major differences in accretion 
geometry and radiation mechanisms. They provide unique applications 
for studying the effects of GR as they allow for more focused 
attempts to model their characteristics.  For example, accretion 
disk spectra from the thermal state are being used to measure 
{\bf black-hole spin} \cite{mcclintock06,liu09}, as are the broad 
iron lines found in the steep power-law state \cite{miller07}.  
This is also where the high-frequency oscillations that are a main 
topic of this white paper are found.  

Simultaneous coverage of outbursts at other wavelengths has led to 
a {\bf unified model for jet formation}  in stellar-mass BHs 
\cite{fender04}. A steady radio jet is observed in the hard spectral 
state, while violent, relativistic jet ejections are observed during 
the transitions from the spectrally hard to spectrally soft states. 
Radio/X-ray luminosity relations initially observed in stellar-mass 
BHs \cite{gallo03} have now been extended to include AGN 
\cite{merloni03,falcke04}, suggesting a scale invariance of the 
jet-accretion coupling in accreting black holes .

\begin{figure}[t]
\hspace{-1.5cm}
\psfig{figure=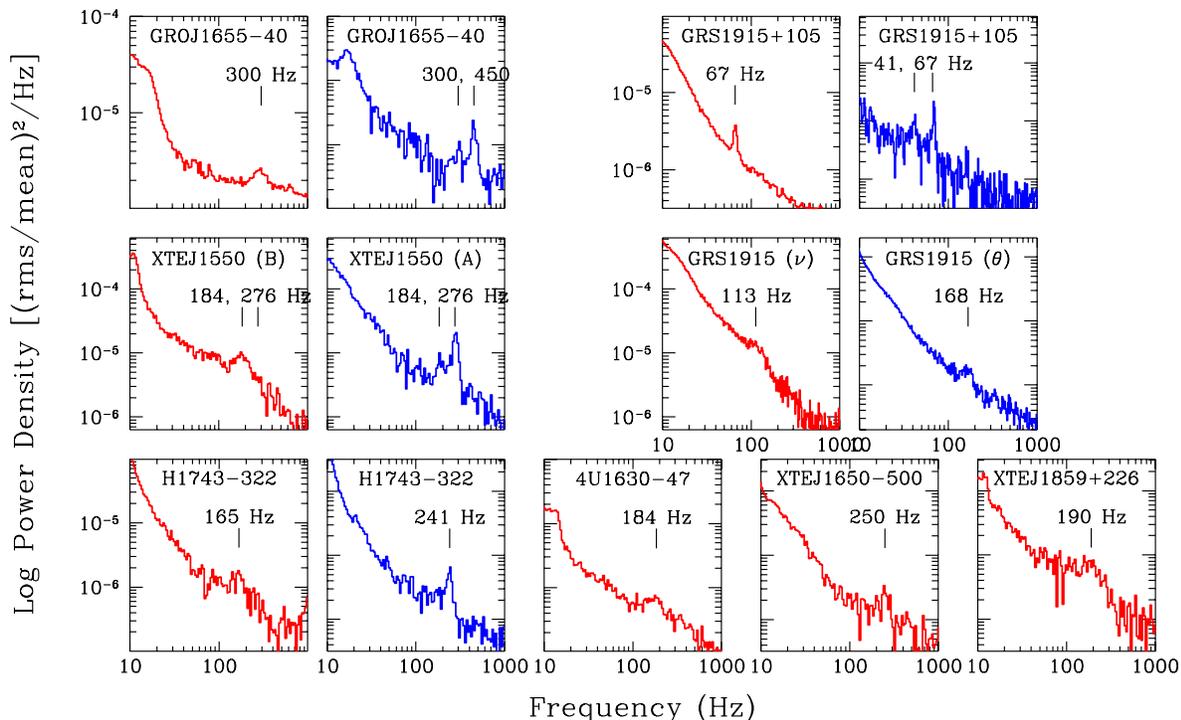,height=10.1cm,angle=-90}
\vspace{-1cm}
\caption{\footnotesize HFQPOs detected in the X-ray PDS of 
BH and BHC systems.  This figure shows the entire sample of 
HFQPOs detected at high significance.  Blue traces are used 
for PDS for 13--40 keV.  Red traces show PDS for a broader 
energy range, either 2--30 or 6--30 keV \cite{rm06}.}
\label{fig:hfqpos}
\end{figure}

{\bf High-frequency QPOs} ($\nu>30$ Hz) have been discovered in 7
BH transients (see Figure~\ref{fig:hfqpos}), five of these in the 
last decade. In 4 of those sources, pairs of HFQPOs have been 
observed, with the frequencies being consistent with a 3:2 ratio. 
While the single peaks have been observed to drift in frequency by 
up to 15\%, the frequencies of the pairs are very stable on time 
scales of years.  There are indications \cite{rm06} that the 
frequencies of the pairs scale inversely with the mass of the black 
hole, as determined from dynamical measurements of the binary 
companion. This scaling, combined with the frequencies and stability, 
suggest that the QPO pairs are rooted in GR, and the properties of
HFQPOs have sparked theoretical interest, leading to models that 
incorporate GR phenomena (e.g., \cite{torok05,sr06}).  While neutron 
star systems also exhibit QPOs, their properties are clearly different 
from the BH HFQPOs.  We expect the BH systems to be cleaner, without
intrinsic magnetic fields or solid surfaces, so the stable frequencies
of HFQPOs may be more likely to be fundamentally related to strong 
gravity.

While HFQPOs are restricted to a narrow range of X-ray properties, 
{\bf low-frequency QPOs} ($<$30 Hz) are found at some level in all 
BH states.  The most common type of LFQPO can be followed in frequency 
from the hardest spectral state to the softest spectral state as it 
increases from $\sim$0.1~Hz up to $\sim$10~Hz, tracing changes in 
the size of the inner disk or Comptonizing corona.  Phase-resolved 
spectroscopy reveals rapid changes in the strength of the iron line 
\cite{mh05}, which has been modeled in terms of a tilted accretion 
disk in strong gravity (e.g., \cite{schnittman06,fb08}).  

\section{X-ray Timing and Black Hole Astrophysics in the Next Decade}

In the next decade, we must accelerate progress for quantitative
applications with GR in investigations of accreting BHs.  The
radiation properties of matter in strong gravity can be used to
conduct fundamental investigations in physics, while constraining the
physical properties of the BHs, i.e., mass ($M$) and spin\footnote{The 
dimensionless spin parameter is $a_{*} = c J / G M^2$, where $J$ is the 
BH angular momentum, $c$ is the speed of light, $G$ is the gravitational 
constant, and the value of $a_{*}$ lies between 0 and 1.}.  Modeling
relativistic accretion disks and broad iron K$\alpha$ emission lines
in X-ray spectra are current techniques for measuring BH parameters.
We must continue to develop these approaches, as well as engaging in
additional techniques, such as interpretations of QPOs and X-ray 
polarimetry.  For all of the models used to interpret the data, there 
are concerns about assumptions and systematic uncertainty.  Thus, it 
is necessary to use multiple techniques and to obtain measurements of 
BH parameters whenever possible.  This motivates the strategy to 
{\bf focus on BH spin} \cite{rm06}, and to capitalize on the continuing 
enterprises that seek BH mass measurements.

The primary focus of this white paper is the opportunity for advancement 
related to X-ray timing measurements, and the HFQPOs represent one of the 
most enticing opportunities.  HFQPOs potentially offer the most accurate 
constraints on BH mass and spin, since the frequencies are measured in 
a straightforward manner, with no distortions due to distance, reddening, 
or other effects common in astrophysics.  The expected link between 
{\bf HFQPOs and strong gravity} is based on two arguments. First, there 
is the relationship between frequency and BH mass mentioned above 
(although this is still based on a small number of systems).  Second, 
HFQPO frequencies are as fast as the dynamical frequencies associated 
with the inner edge of the accretion disk.  GR theory imposes an inner 
boundary condition in the form of an innermost stable circular orbit 
(ISCO) that is outside the event horizon.  For example, for the 
cases, $a_{*} =$ \{0.0, 0.5, 1.0\}, the event horizons lie at 
\{2.0, 1.9, 1.0\} $R_{\rm g}$ and the ISCO radii are
\{6.0, 4.2, 1.0\} $R_{\rm g}$, where $R_{\rm g} \equiv G M/c^{2}$.  For a 
given BH with the same illustrative spin values, the maximum orbital 
frequency has values $\nu_{\rm ISCO} = $ \{220, 351, 1615\} Hz 
($M/$10\Msun)$^{-1}$.  GR also predicts that small orbital perturbations 
do not produce closed orbits, and separate oscillation frequencies would 
be observed for the radial ($\nu_R$) and polar ($\nu_\theta$) coordinates 
(e.g., \cite{merloni99}). Both are slower than the Keplerian frequency
($\nu_\phi$), while all three coordinate frequencies depend on both
the BH mass and spin.

\begin{wrapfigure}{l}{3.3in}
\centerline{\includegraphics[height=2.0in]{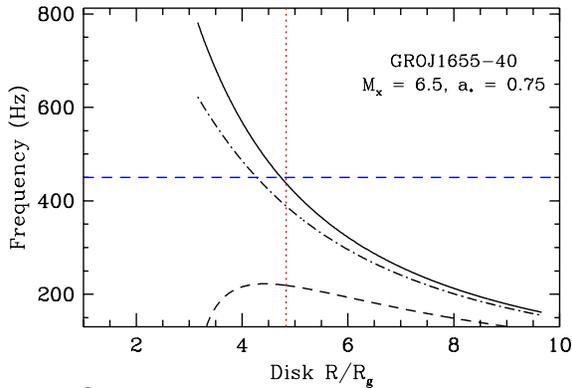}}
\vspace{-0.5cm}
\caption{\footnotesize Frequency map for the inner disk of GRO~J1655--40.  
The black curves show the 3 coordinate frequencies: $\nu_\phi$ (solid),
$\nu_\theta$ (dot-dash), and $\nu_R$ (dashed).  The latter frequency
falls to zero at the ISCO, and for this case $R_{ISCO} = 3.16~R_{\rm
g}$. The vertical, red line shows the location in the disk for maximum
energy release in the Kerr metric ($R_{x} = 4.83~R_{\rm g}$), and this
location happens to be very close to the location where the upper
HFQPO frequency (450 Hz) from this BH intersects the curve for
$\nu_\phi$.\label{fig:bh_nu}}
\end{wrapfigure}

We can now illustrate the expected origin of HFQPOs with the help of
Figure~\ref{fig:bh_nu}. We choose the BH GRO~J1655--40, adopting the
optically determined mass (6.5 \Msun) and the spin value inferred from
the analysis of the thermal state X-ray spectra ($a_* \sim 0.75$).
The curves for the coordinate frequencies ($\nu_\phi$, $\nu_\theta$, 
and $\nu_R$) vs. disk radius are shown.  The orbital frequency ($\nu_\phi$) 
at the radius in the disk where the accreting matter would experience
maximum energy loss is 437 Hz, which is remarkably close to one of the 
HFQPO pairs (300 and 450 Hz) actually measured for this X-ray source.
It has also been noted that observed oscillations are unlikely to be 
faster than the dynamical frequency ($\nu_\phi$) at the radius where 
the QPO originates \cite{strohmayer01}. In Figure~\ref{fig:bh_nu}, this 
would imply (for 450 Hz) {\bf $R_{QPO} < 5 R_{\rm g}$}. Furthermore, this 
HFQPO may exclude the possibility of $a_* = 0$ for GRO~J1655--40, since 
in that case (for 6.5 \Msun) $\nu_{max} = \nu_{ISCO} = 338$ Hz.  HFQPOs 
transport us to realms where $R < 10 R_{\rm g}$, and a proper interpretation 
would yield immediate constraints on BH spin when the mass is known.

Thus, the {\bf key questions} for HFQPO investigations are: what
theoretical models can explain these oscillations?  And what new
observations can be made to point us in the right direction? A general
theory of oscillation modes for accretion disks in the Kerr metric 
has been developed \cite{kato01,wagoner99}.  The turnover of the 
$\nu_R$ curve in Figure~\ref{fig:bh_nu} illustrates the natural capacity 
of a relativistic accretion disk to trap and grow high-frequency 
oscillations. Using linear perturbation theory, the normal disk modes 
were calculated, but the predictions are not consistent with the 
3:2 ratio seen for the HFQPO pairs.  It has also been suggested that 
HFQPOs arise from a {\bf non-linear resonance} mechanism \cite{ak01}, 
since resonances are known to exhibit oscillations with commensurate 
frequencies.  Resonances were first discussed in terms of specific 
radii where GR coordinate frequencies scale with a 3:1 or a 3:2 ratio.  
These simple ideas have been replaced by considerations of resonances 
in fluid flow (e.g., \cite{ka05}).  Other models utilize variations 
in the geometry of accretion. In one model, state changes are invoked 
that thicken the disk into an {\bf accretion torus}, where the normal 
modes can yield oscillations with a 3:2 frequency ratio 
\cite{rezzolla03,baf06}.  It is also possible that HFQPOs exhibit 
properties of a {\bf magnetized accretion disk}, e.g., the rotating 
magnetic spiral waves in the accretion ejection instability (AEI) model 
\cite{tv06}.

\begin{figure}
\begin{center}
\begin{tabular}{c}
\includegraphics[height=7cm]{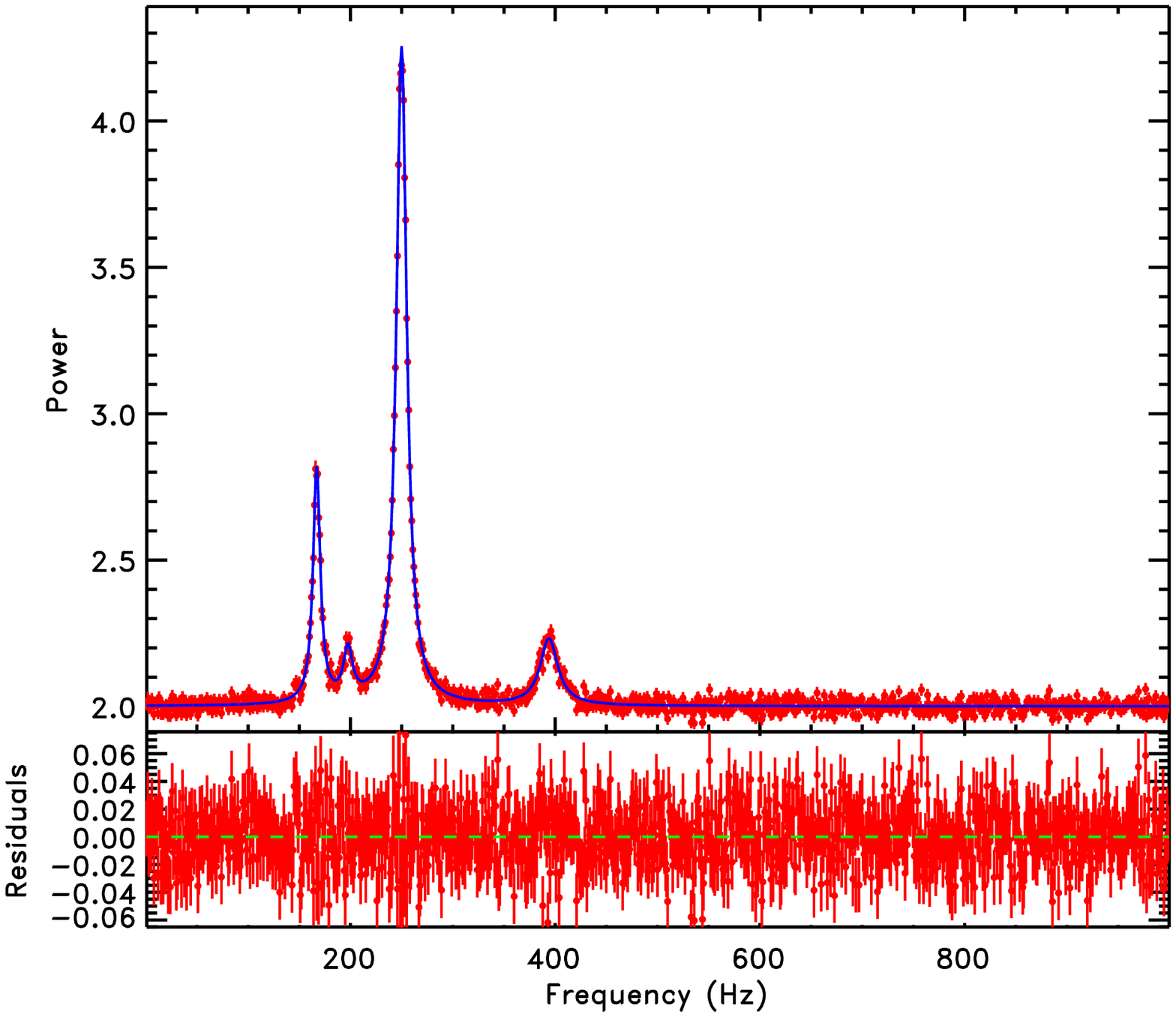}\includegraphics[height=7cm]{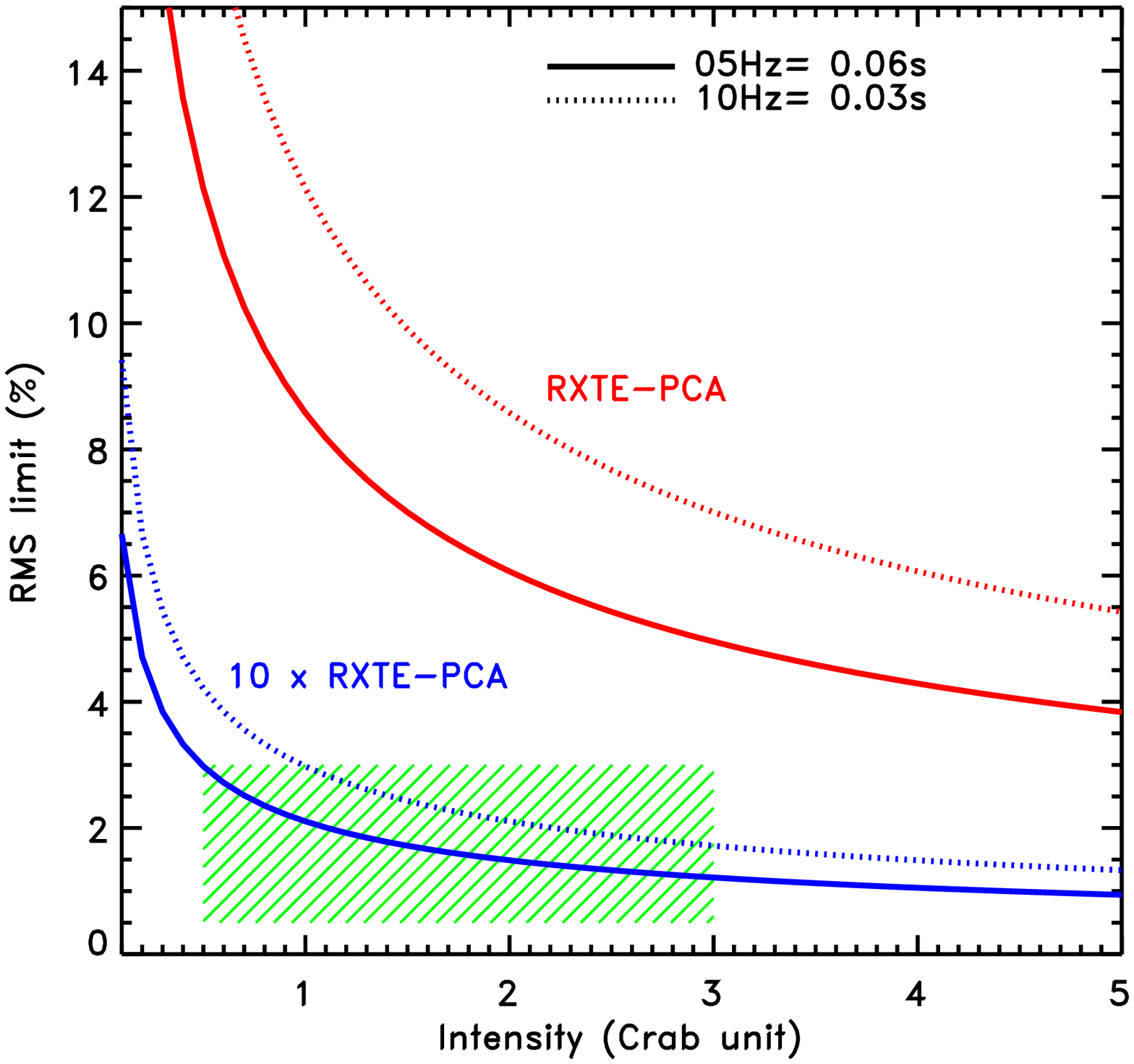}
\end{tabular}
\end{center}
\vspace{-1cm}
\caption[example] 
{\footnotesize {\em Left:} Simulated Power Density Spectrum that could 
be obtained in a 10,000 second exposure by a future X-ray mission with 
10 times the effective area of the {\em RXTE}/PCA.  The QPOs at 250 
and 167~Hz are analogous to HFQPOs at the 3:2 ratios that have been 
previously observed.  This simulated PDS also has weaker QPOs at 400 
and 200~Hz.  Detecting such QPOs with frequencies spaced by integer 
ratios (2:1 in this example) could confirm the non-linear resonance 
mechanism.  {\em Right:} The rms amplitude limit to detect QPOs on 
their coherence time scales ($1/\pi\Delta\nu$) for QPO widths of 
$\Delta\nu = 5$ {\em (solid lines)} and 10 Hz {\em (dashed lines)}, 
as a function of source count rate normalized to the intensity of the 
Crab.  The region where BH QPOs fall is shown as a green-dashed 
rectangle. The PCA limits are shown in {\em red}, and the limit 
that would be obtained with 10 times more effective area is shown 
in {\em blue}.\label{fig:10x}}
\end{figure}

Both the observational and theoretical sides of the HFQPO question
suffer from a fundamental problem.  The observations of HFQPOs in BHs
suffer from chronic signal-to-noise starvation.  Half of the
detections (see Figure~\ref{fig:hfqpos}) were only achieved after 
averaging PDSs over several observations.  The effort to link 
oscillations to theory requires a more complete representation of 
the harmonic structure of HFQPOs that can be compared to predictions 
derived from full modal analyses.  The {\bf current observational 
deficiencies clearly require new missions with larger collecting area}. 
As illustrated in Figure~\ref{fig:10x}, a factor of 10 in area would 
open a new window in the study of millisecond oscillations from 
accreting systems.  In addition, to the possibility of detecting
other resonances, detection of the oscillations within time intervals 
over which they are coherent would permit identification of the 
mechanism of decoherence and should provide insight to their physical 
origin. This is a key step in understanding the QPOs so that they 
can be exploited as tools to study fundamental physics \cite{Kaaret04}.
On the theoretical side, {\bf current models are deficient in terms of
specifying the radiation mechanisms} that would imprint a given
oscillation mode into the X-ray light curve. This step is critically
important, given the fact that HFQPOs and LFQPOs are commonly tied to 
a hard X-ray component, rather than the thermal component that can be 
directly attributed to the accretion disk. Such considerations apply 
to both analytical models and to GR-MHD simulations (see \cite{noble09}).

Observations of LFQPOs with a large effective area X-ray timing mission 
also provide a major opportunity for the next decade.  The evolution 
of LFQPO properties as a source moves from one state to another can 
provide unique insights in the underlying changes in the accretion 
flow geometry. Moreover, LFQPOs probably also originate close enough 
to the BH for GR to have a significant effect on the QPO formation.  
One of the models for LFQPOs in BHs is Lense-Thirring precession
of the inner part of the disk, and if this is correct, then this 
would provide yet another independent measure of BH spin. Two areas of 
LFQPO research that can be opened up further by next generation timing 
instrument are QPO folding (e.g., \cite{tk01}) and phase resolved 
spectroscopy. These techniques are intrinsically better suited for 
LFQPOs than HFQPOs, because of their lower frequencies and higher
amplitudes, but require count rates higher than typically obtained 
with current instrumentation.  While crude attempts have been made 
to measure the energy spectrum of QPOs \cite{gilfanov03,mh05}, 
{\bf phase resolved spectroscopy}, in which spectral changes are 
followed throughout a QPO cycle, will provide more direct information 
on the connection between the QPO and the individual spectral 
component, some of which provide their own spin constraints. 

While there are other more advanced time-domain analysis techniques
(e.g., phase lags, coherence, etc.) that would benefit from a large 
effective area X-ray timing mission in the next decade, we close 
this discussion by emphasizing the link between BH jets and X-ray 
timing signals on both long and short time scales.  Characterizing
the connection between jets and spectral states discussed above has 
depended on multi-wavelength observations, including $\sim$daily 
monitoring of BH transients as they move from state-to-state
over periods of months (e.g., \cite{kalemci05}).  Independently
of X-ray observations, we are expecting two major improvements
to these studies over the next several years.  One of these is
the great increase in radio capabilities expected with facilities
that provide all-sky monitoring like the Low Frequency Array
(LOFAR) and the Murchison Widefield Array (MWA).  With such
capabilities, we will obtain radio data while interesting BH
behavior is occurring rather than observing after such behavior
has occurred.  Secondly, there have been advances in theoretical
modeling of the emission from jets (e.g., \cite{mnw05}), allowing
us to connect measurements to physical jet properties.  However, 
to take advantage of these improvements requires that we also 
keep up our ability to monitor these sources in the X-ray band.

\section{Observational Requirements for X-ray Timing}

The state of the art in X-ray timing is the {\it Rossi X-ray Timing
Explorer (RXTE)} mission.  {\em RXTE} has an all-sky monitor that has 
been essential in tracking the behavior of accreting X-ray sources,
but provides little detailed timing information, and a 
narrow-field/large-area detector array, the Proportional Counter Array
(PCA), that produced most of the scientific advances described above.  
The only mission currently scheduled for launch that will build on 
{\em RXTE}'s timing capabilities is the Indian {\em ASTROSAT} mission.  
{\em ASTROSAT} will contain an X-ray detector array comparable to the 
PCA, but will not offer any significant advance beyond the PCA in 
sensitivity or energy resolution.

The scientific goals described above require new instrumentation.  
The key observational goals are increased sensitivity (both detection 
of weaker QPOs and detection of known QPOs on shorter time scales), 
improved energy resolution, and a capability for monitoring (since 
the BH sources are highly variable and certain signals, such as
HFQPOs, only occur in certain spectral states).  For bright sources, 
defined as sources for which the source counting rate is much larger 
than the background counting rate, the time for detection of a QPO 
signal varies as $T \propto 1/A^2$, where $A$ is the effective 
area \cite{Kaaret04}.  An increase by a factor of 10 in $A$ will lead 
to a decrease of two orders of magnitude in the time required for the 
detection of QPO signals.  This would enable a new X-ray timing mission 
to achieve a qualitative advance in the measurement of QPOs by 
permitting the detection of kHz QPOs within their coherence time
(see Figure~\ref{fig:10x}). 

Non-focusing detector arrays are effective timing instruments for the 
brightest sources.  Next generation detector arrays are likely to be 
based on solid state rather than gas detectors that offer improved 
response at high energies and improved energy resolution relative to 
the PCA \cite{Kaaret01,Chakrabarty08}.  For weaker sources, sheer area 
is not the only concern, and the background counting rate must also be 
considered.  A disadvantage of large, non-focusing detector arrays is 
their high background counting rates.  This limits the effectiveness 
of such arrays for weak sources for which focusing telescopes are 
strongly preferable.  Options include the use of a single detector at 
the focus of a large telescope \cite{Barret08,Elvis04} or an array of 
small telescopes each with a compact detector \cite{Gorenstein04}.  
Focusing telescopes can achieve very large areas at low energies, 
below $\sim$10~keV.  The small detector sizes for focusing telescopes 
also permit much improved energy resolution relative to non-focusing 
detector arrays.

\section{Summary and Conclusions}

Accreting black hole systems provide us with a unique probe of strong
gravity, allowing us to address several important questions.  Are the 
properties of the inner regions of the accretion disk consistent with
the predictions of GR?  What is the radius of the ISCO?  What are the
spin rates of BHs, and does precession of the inner disk around a 
rapidly rotating BH occur?  In addition, BH studies address questions 
of the properties and production of relativistic jets.  To date, the 
work that has been done by X-ray timing has set the stage for producing 
reliable measurements of BH spin and testing GR in the next decade.
Here, we have especially emphasized the opportunity presented by the 
studies of HFQPOs.  These weak signals are close to the {\em RXTE} 
detection limit, and larger effective area missions are required to 
use HFQPOs as tools to study fundamental physics.  

\begin{multicols}{2}

\end{multicols}


\begin{thebibliography}{}
\footnotesize
\setlength{\itemsep}{0em}
\setlength{\parsep}{0pt}
\setlength{\baselineskip}{0pt}
\renewcommand{\baselinestretch}{0.4}

\bibitem{ak01} Abramowicz, M.A. \& Klu{\'z}niak W., 2001, A\&A, 374, L19

\bibitem{Barret08} Barret, D., 2008, Proc.~SPIE, 7011, 10

\bibitem{baf06} Blaes, O.~M., Arras, P., \& Fragile, P.~C., 2006, MNRAS, 369, 1235

\bibitem{blandford00} Blandford, R., et al., 2000 Decadal Review, 
Panel Report \#1

\bibitem{casella08} Casella, P., et al., 2008, MNRAS, 387, 1707

\bibitem{Chakrabarty08} Chakrabarty, D., Ray, P.S., Strohmayer, T.E.,
2008, AIP Conf.~Proc., 1068, 227

\bibitem{Elvis04} Elvis, M., 2004, AIP Conf.~Proc., 714, 459

\bibitem{falcke04} Falcke, H., et al., 2004, A\&A, 414, 895

\bibitem{fender04} Fender, R.P., et al., 2004, MNRAS, 355, 1105

\bibitem{fb08} Fragile, P.C. \& Blaes, O.M., 2008, ApJ, 687, 757

\bibitem{gallo03} Gallo, E., et al., 2003, MNRAS, 344, 60

\bibitem{gilfanov03} Gilfanov, M., et al., 2003, A\&A, 410, 217

\bibitem{Gorenstein04} Gorenstein, P., 2004, AIP Conf.~Proc., 714, 431

\bibitem{Kaaret01} Kaaret, P. et al., 2001, AIP Conf.~Proc., 599, 678

\bibitem{Kaaret04} Kaaret, P., 2004, AIP Conf.~Proc., 714, 423

\bibitem{kalemci05} Kalemci, E., et al., 2005, ApJ, 622, 508

\bibitem{kato01} Kato. S., 2001, PASJ, 53, L37 

\bibitem{ka05} Klu{\'z}niak, W., \& Abramowicz, M.~A., 2005, Ap\&SS, 300, 143

\bibitem{liu09} Liu, L., et al., 2009, ApJ, 691, 847

\bibitem{mnw05} Markoff, S., Nowak, M.A., \& Wilms, J., 2005, ApJ, 635, 1023

\bibitem{mcclintock06} McClintock, J.E., et al., 2006, ApJ, 652, 518

\bibitem{mr06} McClintock, J.E. \& Remillard, R.A. 2006, in ``Compact 
Stellar X-ray Sources'', eds. W.~G.~H. Lewin \& M. van der Klis, Cambridge 
University Press, 157

\bibitem{mchardy06} McHardy, I.M., et al., 2006, Nature, 444, 730

\bibitem{merloni99} Merloni, A., et al., 1999, MNRAS, 304, 155

\bibitem{merloni03} Merloni, A., et al., 2003, MNRAS, 345, 1057

\bibitem{miller07} Miller, J.M., 2007, ARA\&A, 45, 441

\bibitem{mh05} Miller, J.M. \& Homan, J., 2005, ApJ, 618, L107

\bibitem{noble09} Noble, S.~C., Krolik, J.~H., \& Hawley, J.~F., 2009, 
ApJ, in press; astro-ph/0808.3140

\bibitem{rm06} Remillard, R.A. \& McClintock, J.E., 2006, ARA\&A, 44, 49

\bibitem{rezzolla03} Rezzolla, L., et al., 2003, MNRAS, 344, L37

\bibitem{sr06} Schnittman, J.D. \& Rezzolla, L., 2006, ApJ, 637, L113

\bibitem{schnittman06} Schnittman, J.D., et al., 2006, ApJ, 642, 420

\bibitem{smith02} Smith, D., et al., 2002, ApJ, 569, 362

\bibitem{strohmayer01} Strohmayer, T., 2001, ApJ, 552, L49 

\bibitem{tv06} Tagger, M. \& Varniere, P., 2006, ApJ, 652, 1457

\bibitem{tk01} Tomsick, J.A. \& Kaaret, P., 2001, ApJ, 548, 401

\bibitem{torok05} T\"or\"ok, G., et al., 2005, A\&A, 436, 1

\bibitem{wagoner99} Wagoner, R.~V., 1999, PhysRpt, 311, 259 

\end{thebibliography}
\end{document}